\documentclass[aps,pra,floatfix,twocolumn,superscriptaddress,10pt]{revtex4-1}
\usepackage{amssymb, amsmath,graphicx,subfigure}
\usepackage{color}
\usepackage{braket}
\usepackage{hyperref}
\usepackage{times}

\begin{document}

\title{High-fidelity initialization a logical qubit with multiple injections}

\author{Zhi-Cheng He}

\affiliation{Key Laboratory of Atomic and Subatomic Structure and Quantum Control (Ministry of Education),\\ 
Guangdong Basic Research Center of Excellence for Structure and Fundamental Interactions of Matter,\\
and School of Physics, South China Normal University, Guangzhou 510006, China}

\author{Zheng-Yuan Xue}\email{zyxue83@163.com}
\affiliation{Key Laboratory of Atomic and Subatomic Structure and Quantum Control (Ministry of Education),\\ 
Guangdong Basic Research Center of Excellence for Structure and Fundamental Interactions of Matter,\\
and School of Physics, South China Normal University, Guangzhou 510006, China}
\affiliation{Guangdong Provincial Key Laboratory of Quantum Engineering and Quantum Materials,\\  Guangdong-Hong Kong Joint Laboratory of Quantum Matter,   and Frontier Research Institute for Physics,\\ South China Normal University, Guangzhou 510006, China}
\affiliation{Hefei National Laboratory,  Hefei 230088, China}

\begin{abstract}
Quantum error correction represents a significant advancement in large-scale quantum computing. However, achieving fault-tolerant implementations of non-Clifford logical gates with reduced overhead remains a challenge in the popular surface code strategy. Recent advances have underscored the need for a substantial code distance to attain complete fault tolerance. Here, we introduce a continuous fault-tolerant scheme for non-Clifford logical gates via multiple injections. Unlike existing protocols that focus on a single logical chain, our approach utilizes multiple logical chains, each can employ the same or different logical rotation angles, to initialize a non-Clifford state. Compared to previous efforts, our protocol significantly alleviates the challenges associated with the requirement for a large code distance and reduces the corresponding resource overhead, making it more feasible to be implemented in current mid-scale chips via the  surface code strategy.

\end{abstract}

\maketitle

\section{Introduction}
Quantum computers promise to surpass their classical counterparts in certain challenging tasks that require exponential speed improvements. However, the fragile nature of quantum information necessitates effective error correction and error mitigation for practical and large-scale quantum computing. As a result, a variety of quantum error correction codes have been developed to safeguard quantum information from the main sources of error \cite{QEC_Shor, QEC_Gottesman, QEC_Steane, QEC_Calderbank, QEC_Laflamme, QEC_Bennett, QEC_Cleve, QEC_Knill}. In quantum error correction codes, a logical qubit is typically encoded by multiple physical qubits, enabling the correction of errors in these physical units. Among these codes, the stabilizer code stands out for their ability to transform various quantum errors into Pauli errors, thereby simplifying the process of error detection and correction. The surface code, a prominent example of the stabilizer code, utilizes topology-based techniques to encode information within a two-dimensional lattice structure through local interactions among neighboring qubits. This method, recognized for its experimental viability, has attracted considerable attention in practical applications, establishing itself as a leading candidate for implementing future large-scale quantum computers \cite{Surface_code, Surface_code_Kitaev, Surface_code_Raussendorf, Surface_code_high_threshold, Surface_code_error_rate_1, Surface_code_google, Surface_code_real_noise, Surface_code_repeated, Superconducting_to_surface_code, Surface_code_classical_processing}.

Today, demonstrating the practical applications of a quantum computer has been a critical goal in the field of quantum computation \cite{arute2019, simon1997, bernstein1997, lloyd1996, aaronson2013}. To this end, several promising algorithms have been proposed, including factoring large integers, searching algorithms, variational quantum algorithms, quantum approximate optimization algorithms, and the quantum phase estimation \cite{shor1994, grover1996, peruzzo2014, cerezo2020, kandala2017, abrams1999}. However, using a surface code or most stabilizer codes, these algorithms require a significant number of non-Clifford quantum gates, which is challenging in practical implementations \cite{Lattice_surgery_Horsman, Lattice_surgery_experiment, Topological_Order_with_a_Twist, Poking_holes_and_cutting_corners, surface_code_with_a_twist, Universal_1, Universal_2}. The primary obstacle arises from the substantial resource overhead associated with implementing a non-Clifford logical gate. Specifically, obtaining non-Clifford logic gates typically involves fault-tolerant initialization of an auxiliary logical qubit into a non-Clifford state via the magic state distillation technique
\cite{Magic_state_distillation_low_overhead, Magic_state_distillation_not_costly}.
This process is resource-intensive and significantly contributes to the overall resource overhead of the surface code-based quantum computation. Consequently, exploring alternative methods to circumvent magic state distillation for implementing logical gates is a critical task.

To achieve logical rotation without relying on magic state distillation, small-angle rotations are of primary interest, as they have been extensively utilized in various quantum algorithms  
\cite{Trotterization_1, Trotterization_2, VQEs_1, VQEs_2, Nielsen1}. In Trotter simulation, it utilizes a quantum circuit comprising a large number of Clifford gates and rotation gates with very small angles. For variational quantum algorithms, it needs minor adjustments according to the output of the previous algorithm's cycle. Recently, protocols based on single-qubit operation and post-selection sparked considerable interest \cite{Analog_Rotations, Lis_paper, Preparation_ZKD, Transversal_Injection, post_selection, post_selection_Fujii_1, post_selection_Fujii_2}. Among them, The protocols enable the preparation of certain special non-Clifford states without the need for magic state distillation, provided that the resulting state is obtained by rotating a Clifford state by a sufficiently small angle \cite{post_selection, post_selection_Fujii_1, post_selection_Fujii_2}. The protocol begins with a fault-tolerantly initialized Clifford state, followed by the application of Z- or ZZ-rotations to initialize non-Clifford states. Although some physical Z errors may pass error detection and lead to an incorrect logical state, the logical subspace remains unchanged, only the superposition coefficient of the logical state is modified.
Furthermore, it has been found that an equivalent form of fault tolerance can be achieved against these single-qubit gate errors, referred to as "continuous fault tolerance". This implies that the infidelity of the initialized logical qubit with a small rotation angle is rapidly suppressed as the code distance increases. With this promise of fault tolerance, the resource overhead for many quantum algorithms that frequently utilize small-angle Z-rotations can be significantly reduced. However, achieving continuous fault tolerance requires a relatively large encoding distance, which incurs substantial space costs. This requirement presents a significant challenge, as expanding the code array is difficult nowadays.

Here, we propose a multiple injection protocol to achieve continuous fault tolerance. Unlike other initialization protocols that focus on injecting into a single logical chain, our approach employs multiple logical chains with the same or different injections to initialize non-Clifford states. Compared to previous protocols, our method significantly alleviates the requirement for a large code distance to achieve continuous fault tolerance, effectively transforming space costs into time costs, thereby making it feasible for implementation in mid-scale surface codes. Besides, the overall resource overhead of our scheme is reduced as it can achieve continuous fault tolerance via smaller code distance and lower overhead. Our result can be immediately applied to current experimental setups to obtain high-fidelity logical states within the surface code framework.

\section{The multiple injection protocol}

In this section, we review the single injection scheme and present our multiple injection protocol to initialize a non-Clifford logical state.

\subsection{Injection protocols}
First, we consider a surface code that has been initialized into an eigenstate of the logical operator $X$, i.e., $\ket{+}_{L}$, in a fault-tolerant way \cite{Preparation_ZKD}. For the single chain injection scheme\cite{post_selection,post_selection_Fujii_1}, one implements physical single-qubit rotation gates around the Pauli $Z$ axis to a set of qubits $Q$ along a chain, i.e.,
\begin{eqnarray}
\label{R_z to Q}
\prod_{j\in Q} R_{z}(\theta_{p_j})=\prod_{j\in Q}\left(\cos{\frac{\theta_{p_j}
}{2}}\cdot I+\text{i}\sin{\frac{\theta_{p_j}}{2}}\cdot Z\right),
\end{eqnarray}
where $\theta_{p_i}$ labels the angle of rotation around the $Z$ axis for the $i$th physical qubit. In general cases, these $Z$-rotations just induce single-qubit gate errors for a surface code. However, if the injected qubits form a logical chain $C$, these rotation gates may induce logical rotations of
\begin{eqnarray}
\label{R_z to Q 2}
\prod_{j\in C} R_{z}(\theta_{p_j})
&=&\sqrt{P}\left(\cos{\frac{\theta_{c}}{2}}\cdot I
+\text{i}\sin{\frac{\theta_{c}}{2}}\cdot Z_{L}\right)+\text{others} \nonumber\\
&=& \sqrt{P} R_{Z,L}(\theta_{c}) +\text{others},
\end{eqnarray}
where $Z_{L}=\prod_{j\in c} Z_{j}$, labels the logical $Z$ operator, $R_{Z,L}$ labels the logical $Z$-rotation, and $\theta_{c}$ labels the rotation angle that this logical chain contributes to the angle of logical rotation. In the case of single injection, only one chain is operated, thus the rotation angle for the logical $Z$ operation is equal to the angle of one logical chain, i.e., $\theta_{L}=\theta_c$. Here, we temporarily assume $\text{i}^d = \text{i}$ for obtaining Eq. (\ref{R_z to Q 2}), and for general logical operations, see Appendix A. The other terms in Eq. (\ref{R_z to Q 2}) can be discarded by measuring the stabilizers. The post-selecting process will ensure the correct trajectory of the measurement values. It contributes in two key aspects. Firstly, the majority of detectable errors typically occur during the single qubit Z-rotation and stabilizer measurement cycle, leading to difference in the stabilizer trajectory. Post-selection ensures that the measurement trajectory remains unchanged, thereby mitigating the effects of these errors. Secondly, by successfully passing the post-selection with a probability $P_c$, the other terms in Eq. (\ref{R_z to Q 2}) can be disregarded. Consequently, the final logical state is guaranteed to be correct. The success probability of post-selection can be easily obtained by calculating the normalization factor as
\begin{eqnarray}
\label{P of post-selection}
P_c = \prod_{j\in Q} \sin^{2}{\frac{\theta_{p_j}}{2}}+\prod_{j\in Q} \cos^{2}{\frac{\theta_{p_j}}{2}}.
\end{eqnarray}
And the angle for the logical chain is
\begin{eqnarray}
\label{Angle of logcical}
\theta_{c}=\pm 2\arcsin{\left[\frac{\prod_{j\in Q} \sin{\frac{\theta_{p_i}}{2}}}{\sqrt{P_c}}\right]},
\end{eqnarray}
where the positive or negative sign depends on the trajectory of stabilizer measurement. Generally, it depends on the binary sum of the value of Z-stabilizers near the logical chain. If the binary sum of the value is zero (one), we obtain a positive (negative) value for the logical rotation angle.

Next, we consider the multiple chain injection scheme. In this case, the set of qubits $Q$ consists of several subsets, which are logical chains, i.e., $Q=\sum_{n} C_{n}$, where $C_{n}$ labels the $n$th logical $Z$ chain. And, the Eq.(\ref{R_z to Q 2}) will change to
\begin{eqnarray}
\label{R_z to Q_s}
\prod_{j\in Q} R_{z}(\theta_{p_j})
&=&\prod_{n}\sqrt{P^{n}_c}R^{n}_{Z,L}(\theta^{n}_{c})+\text{others} \notag\\
&=& \prod_{n}\sqrt{P^n_c} R_{Z,L}\left(\sum_n\theta^{n}_{c}\right) + \text{others},
\end{eqnarray}
where $P^{n}_c$ and $R^{n}_{Z,L}$ label the success probability and the contribution to logical rotation angle, corresponded to $n$th logical chain. Similar to the previous single chain injection scheme, the post-selection here is also applied to ensure that the obtained trajectory is the same as that of $\ket{+}_{L}$. It is worth noting that the operation cannot be implemented to two nearby chains, which will induce the collapsed logical states to unwanted results. This feature will result in the number of chains being limited to a maximum of $n=(d+1)/2$. After that, the final logical rotation angle becomes $\theta_{L}=\sum_{n}\theta^{n}_{c}$. Setting all the $\theta^{n}_{c}=\theta_{c}$ for simplicity, the implemented logical rotation will reduce to $\theta_{L}=n\theta_{c}$. However, when the quantities of corresponding chains with positive rotation angle are approximately equal to the chains with negative rotation angle, the final logical operator will nearly form an identity. To avoid this worst situation, we can rotate the single-qubit for negative angles if the trajectory suggests a negative chain. For instance, if the stabilizer trajectory indicates that most of the $\theta^{n}_{c}$ have positive values but one has a negative value, we perform $Z$-rotations with a negative angle on the physical qubits in that specific chain. Consequently, the rotation angle becomes the desired positive value. Thus, we can obtain desired logical Z-rotations whatever the trajectory reports.

\begin{figure}[tbp]
    \centering \includegraphics[width=1.0\linewidth]{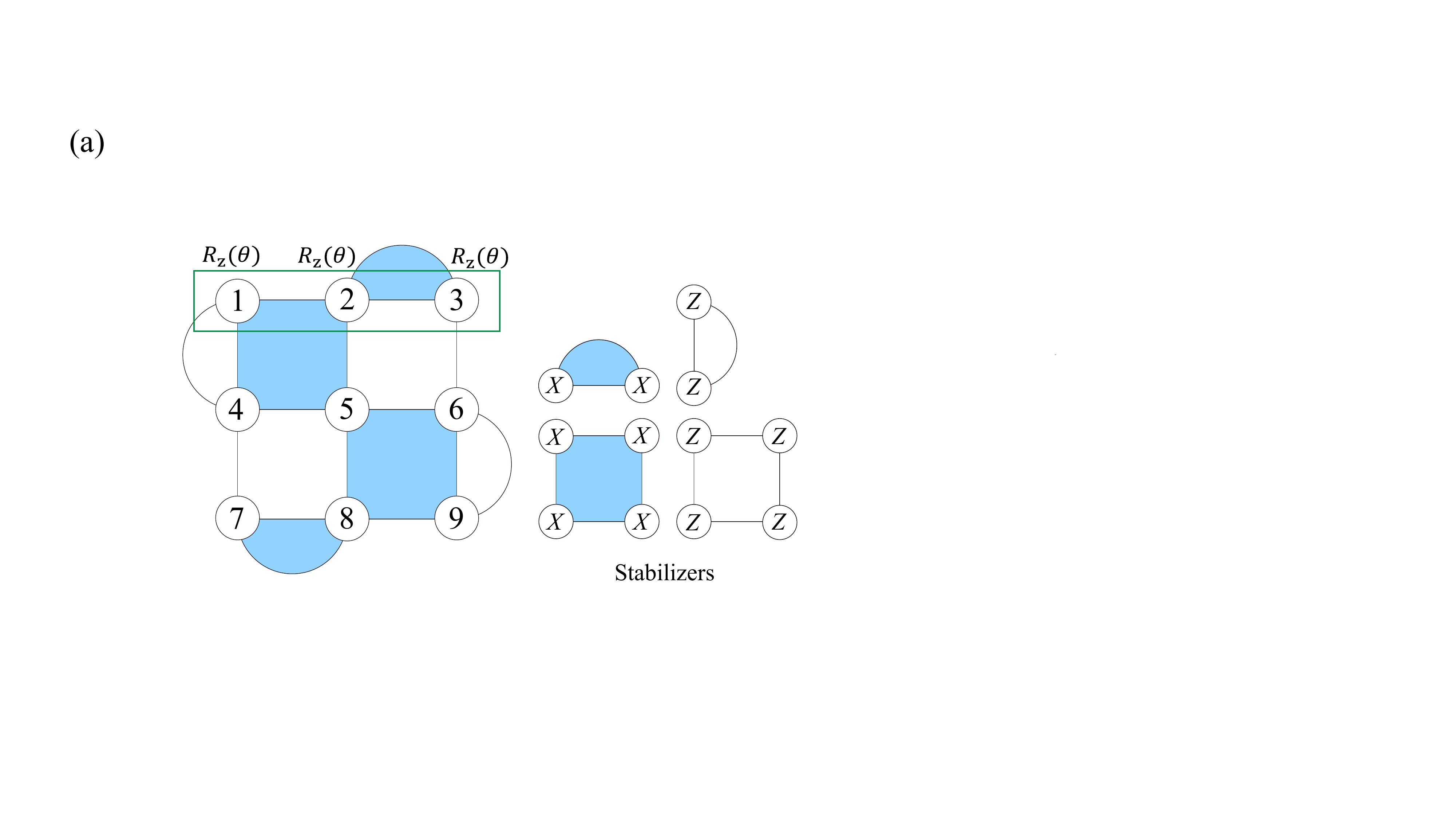}
\caption{The illustration of the injection protocol, with a distance 3 rotated surface code as a typical example. The blue (white) tiles represent the $X$ ($Z$) -type stabilizers. The logical operator $X_{L}$ respect to this layout has three forms, i.e., $X_{L}=X_{1}X_{4}X_{7}$, $X_{L}=X_{2}X_{5}X_{8}$ and $X_{L}=X_{3}X_{6}X_{9}$. The logical operator $Z_{L}$ is similar,  $Z_{L}=Z_{1}Z_{2}Z_{3}$, $Z_{L}=Z_{4}Z_{5}Z_{6}$ and $Z_{L}=Z_{7}Z_{8}Z_{9}$. When initializing a logical state by using the single (multiple) injection protocol, the single-qubit $Z$-rotations will be implemented to the physical qubits, which  belong to one (some) of logical chain.}
\label{Lattice_and_circuit}
\end{figure}

\subsection{Examples for the d=3 surface code}

To illustrate the initialization protocol effectively, we consider a distance 3 surface code as a representative example for creating a non-Clifford logical state. We begin by a logical state that has been prepared in $\ket{+}_L$ \cite{Preparation_ZKD}, where the measurement  of stabilizers yield a trajectory of $\{0MM00MM0\}$. The values of the trajectory correspond to the stabilizers $X_{2}X_{3}$, $X_{1}X_{2}X_{4}X_{5}$, $X_{5}X_{6}X_{8}X_{9}$, $X_{7}X_{8}$, $Z_{1}Z_{4}$, $Z_{2}Z_{3}Z_{5}Z_{6}$, $Z_{4}Z_{5}Z_{7}Z_{8}$, and $Z_{6}Z_{9}$ in Fig. \ref{Lattice_and_circuit}.
When we obtain a trajectory of $\{0000 0110\}$ during the initialization of $\ket{+}_L$, we first select two non-adjacent logical chains, $C_{1}=Z_{1}Z_{2}Z_{3}$ and $C_{2}=Z_{7}Z_{8}Z_{9}$, to apply single-qubit Z-rotations with same angles, $\theta_p$. The operation becomes
\begin{eqnarray}
\label{EXAM R_z to Q}
\prod_{j\in Q} R_{z}(\theta_{p})
&=&\sqrt{P^{1}_{c}}\sqrt{P^{2}_{c}} R^{1}_{Z,L}(\theta_{c})R^{2}_{Z,L}(\theta_{c}) + \text{others}.\notag\\
&=& P_c R_{Z,L}(2\theta_{c}) + \text{others}.
\end{eqnarray}
After this step, the final logical rotation angle yields $R_{Z,L}(2\theta_c)$ upon successful post-selection. Notice we performed the physical single-qubit Z-rotations with the same rotation angle $\theta_p$, due to the trajectory indicates that two logical rotation angles are negative. Specifically, the logical chain $C_{1}=Z_{1}Z_{2}Z_{3}$ is adjacent to two Z-type stabilizers, $Z_{1}Z_{4}$ and $Z_{2}Z_{3}Z_{5}Z_{6}$. The binary sum of these two Z-type stabilizer values imply a negative logical angle, in Eq. (\ref{Angle of logcical}). Similarly, the logical chain $C_{2}$ exhibits a similar situation to that of $C_{1}$.

We then initiate a new stabilizer measurement cycle, which yields a new trajectory, upon the measurement outcomes. The post-selection is employed to guarantee that the new trajectory is identical to the original trajectory. If the two trajectories differ, the logical state would be projected into an unintended logical subspace in Eq. (\ref{EXAM R_z to Q}). If the two trajectories are the same, we successfully pass the post-selection, and obtain a non-Clifford logical state $R_{Z,L}(2\theta_c)\ket{+}_L$ consequently.

\section{Discussions on the gate performance}

In this section, we present the performance of our protocol, showcase it advantages compared to single injection schemes.

\subsection{The Continuous fault-tolerance}

We now explore the continuous fault-tolerance in both single and multiple injections, considering a general Pauli circuit noise model. The circuit noise model is employed to represent the errors introduced by Pauli gates. Specifically, using the depolarizing channel, a single-qubit error is described as one of depolarizing gates $\{X, Y, Z\}$ and its occurring probability is called physical error rate. Moreover, a two-qubit error depolarizing gate is modeled as one of two-qubit gates $\{I, X, Y, Z\}^{\otimes 2}$. If an error occurs, the altered operation is described as a consisted operation with a desired perfect operation and an added depolarizing gate. The vulnerable operations include reset operation, single-qubit gates, two-qubit gates and measurements.

Both single and multiple injection protocols can tolerate main circuit errors through the implementation of two stabilizer cycles. These errors can alter the measurement results, leading to discrepancies between successive measurements. Upon detection of such differences, the prepared logical state is discarded. Nevertheless, there has a particular type of error that can affect the logical state without changing the measurement outcome of stabilizers. Specifically, this is the $Z$ error that occur before or after the single-qubit Z-rotations, which affects the superposition coefficient of the logical state rather than altering the logical subspace itself. This is because a $Z$ error is equivalent to over rotation, i.e., from the target rotation $R_z(\theta_p)$ to $R_z(\theta_p+\pi)$. Consequently, a single $Z$ error in physical qubits modifies the logical rotation angle to
\begin{eqnarray}
\label{Angle of logcical with error}
\theta_{c}^{e}=\pm 2\arcsin{\left[\frac{\sin{\frac{\theta_{p}+\pi}{2}} \prod_{i\in Q-1} \sin{\frac{\theta_{p}}{2}}}{\sqrt{P}}\right]}.
\end{eqnarray}

For the single injection, the logical rotation angle $\theta_L$ is entirely determined by the rotation angle of the single chain, i.e., $R_{L}(\theta_L)=R_{L}(\theta_c)$.
Therefore, the logical rotation angle is completely modified from $\ket{\psi}_L = R_{L}(\theta_{L})\ket{+}_L$ to $\ket{\psi}^e_{L}=R_{L}(\theta^e_{L})\ket{+}_L$, where $\theta^{e}_{L}=\theta^{e}_{c}$.
It is straightforward to calculate the infidelity between the desired logical state and the erroneous one as
\begin{equation}
\label{Fidelity_single}
F_i =1-\lvert \braket{\psi_{e} \lvert \psi}_L\lvert^2
= 1- \frac{1}{2} \left\lvert 1+e^{\text{i}\delta_c}\right\lvert^2,
\end{equation}
where $\delta_c=\theta_c-\theta^e_{c}$ determines the $F_i$. When $\theta_{p}$ is sufficiently small ($\theta_{p} \ll 1$), we can approximate $\theta_c \approx (\theta_{p})^d$, that is, $F_i$ is exponentially suppressed. Since non-Clifford circuits are difficult to simulate on classical computers, we calculate fidelities by utilizing the state of one physical qubit instead of the logical state, see Appendix B for details. We verified our replacement is effective, by compare the result of both one qubit state and the logical state. The fidelity calculation for logical state uses simulated surface codes with a $3\times 3$ qubit lattice and further extend it to a $3\times 5$ case, through Qiskit\cite{Qiskit}. Notice that the fidelity of the obtained logical state will significantly affect the performance of each single-qubit logical gate. For example, the acceptable $F_i$ for variational quantum algorithms is $10^{-4}$ or even $10^{-6}$.
This requirement can be satisfied by the continuous fault-tolerance property of the single injection protocol, but it requires a larger code distance, which is hardware inefficient and is out of state-of-the-art technology.

\begin{figure}[tbp]
  \centering
  \includegraphics[width=0.9\linewidth]{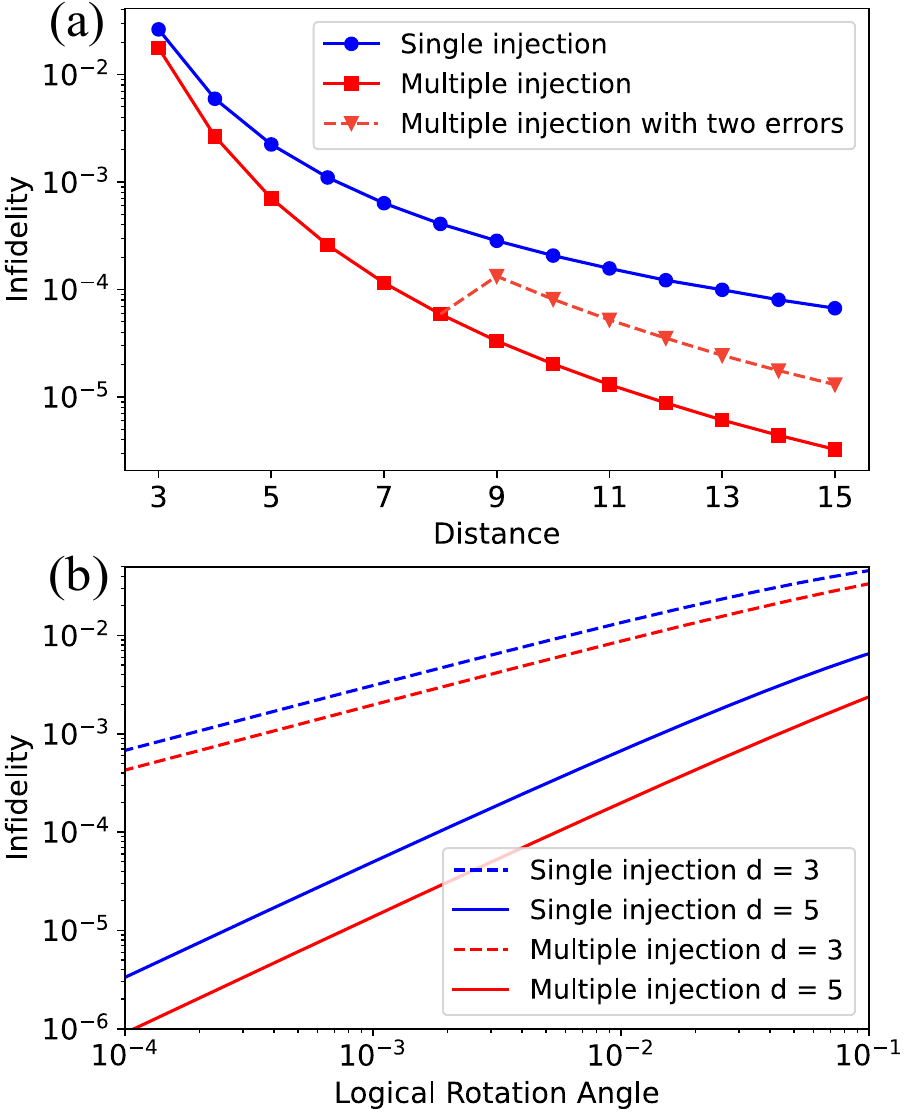}
  \caption{The performance of continuous fault-tolerance with a $Z$ error for both single and multiple injection schemes. (a) The performance for obtaining the logical state $R_{Z,L}(\theta_{L})\ket{+}$ with different code distance when $\theta_{L}=2\pi\times 10^{-2}$. Blue (orange) lines indicates the single (multiple) injection scheme, while the green line indicates a worse situation of two $Z$ errors in the multiple injection scheme. For the multiple injection scheme, $n$ chain injections are used, and thus its infidelity is  quickly reduced with increasing $d$. (b) The performance for obtaining the logical state $R_{Z,L}(\theta_{L})\ket{+}$ with different  $\theta_{L}$. For a same $d$, multiple injection scheme still has a smaller infidelity.}
\label{Infidelity with distance}
\end{figure}

The multiple injection scheme represents a significant advance, as it can greatly reduce the infidelity of the process. Thus, for achieving the required level of infidelity, it needs smaller code distance. Specifically, through multiple injection, the obtained rotation angle for the logical qubit is the cumulative sum of the rotation angle of each chain, that is, $\ket{\psi}_L^{'} = R_{L}(n\theta_{c})\ket{+}_L$, where $n$ denotes the number of chains involved in the operation. Consider that single-qubit Z errors occur in operated chains, the obtained logical rotation will be changed to $\ket{\psi}^{'e}_{L} =R_{L}[(n-n_e)\theta_{c} +n_e\theta^{e}_{c}]\ket{+}_L$, where $n_e$ labels the number of chains with a Z error. Thus, the infidelity for multiple injection scheme is
\begin{equation} \label{Fidelity_multiple}
F_i'=1-\lvert \braket{\psi^{e} \lvert \psi}_L^{'}\lvert^2 
=1- \frac{1}{2}\left \lvert 1+e^{\text{i}\frac{n_e}{n}\delta_c}\right \lvert^2.
\end{equation}
Clearly, with respecting to undetectable $Z$ errors, multiple injection scheme is inherently more robust than the single injection one, as long as $n_e < n$. It is reasonable since the multiple injection scheme employs more data qubits to distill a single logical piece of information. Moreover, the possibility of such multiple errors is negligible small for mid-size qubit-lattice as the physical error rate is currently low enough. The multiple injection scheme results in the same level of infidelity with that of the single injection only when each chain has a $Z$ error, according to Eq. (\ref{Fidelity_multiple}). The probability of this case is an $k$th-order event.
When the code distance is $d=9$, the probability for two errors occurred in the multiple injection scheme, is only about 1/10 of the probability for one error occurred in the single injection scheme, based on the current level of single-qubit gate error rate of $10^{-4}$ \cite{Single_qubit_gate_1,Single_qubit_gate_2}.
Thus, we neglect the events that two or more errors occur when the code distance is small. Specifically, we construct a simple error model where a single-qubit $Z$ error can occur in every physical qubit with the same probability. Then, we assume at most one error can occur in small and middle distance qubit lattice cases ($d\leq 8$) of the multiple-injection scheme, and for all distance of the single-injection scheme. To show a worse situation for our multiple-injection scheme, we also consider the two errors case for larger qubit lattices ($d \geq 9$). Our numerical simulation results for both schemes are shown in Fig. \ref{Infidelity with distance}. Besides, the success probability of the multiple injection scheme increases exponentially with $n$. However, this fact will not significantly impact the success probability for mid-size lattices. As $n$ increases, the rotation angle for each chain decreases since $\theta_{L}=n\theta_{l}^{n}$, which leads to the increase of the success probability for each chain.

\begin{figure}[tbp]
  \centering
  \includegraphics[width=1.0\linewidth]{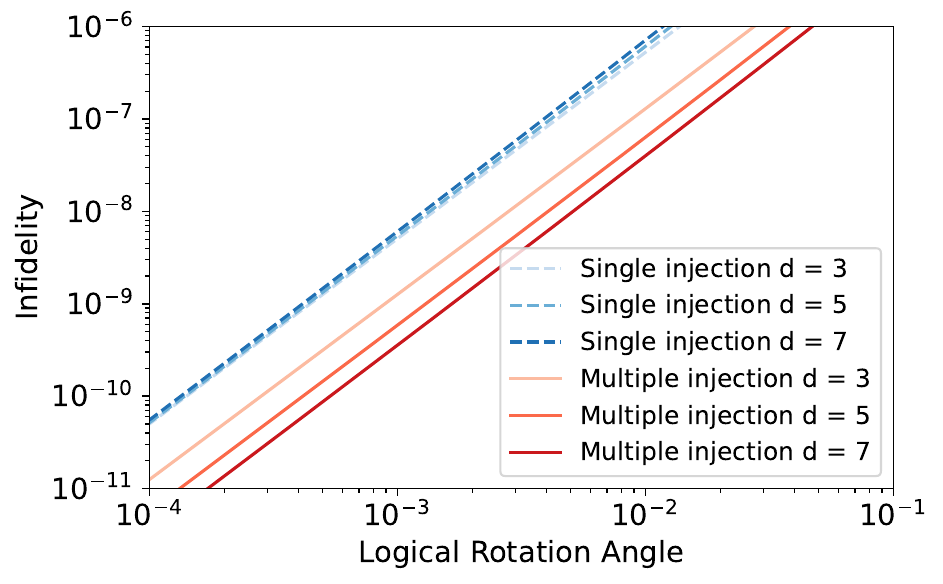}
  \caption{Performance of continuous fault-tolerance with an over rotation error. The performance for obtaining the logical state $R_{Z,L}(\theta_{L})\ket{+}$ with different angle $\theta_{L}$. The error is described as an over rotation $\theta_{e} = (1 + \epsilon)\theta_{p}$, where $\epsilon=0.2$.}
\label{Infidelity with distance_small error angle}
\end{figure}

Furthermore, practically, quantum errors are not simply represented as Pauli operators. When a physical Z-rotation is implemented experimentally, the noise typically causes only a slight difference of the rotating axis or angle. The inaccurate rotation axis can be corrected by the stabilizer measurements. However, the slight difference of the rotating angle can pass the error detection process and changes the target rotation from $\theta_{p}$ to $\theta_{e} = \theta_{p} + \epsilon$, where $\epsilon \gg \theta_{p}$. We demonstrate that the multiple injection scheme also performs better than the small rotation angles under realistic noise conditions, as shown in Fig. \ref{Infidelity with distance_small error angle}.

\begin{figure}[tbp]
  \centering
\includegraphics[width=\linewidth]{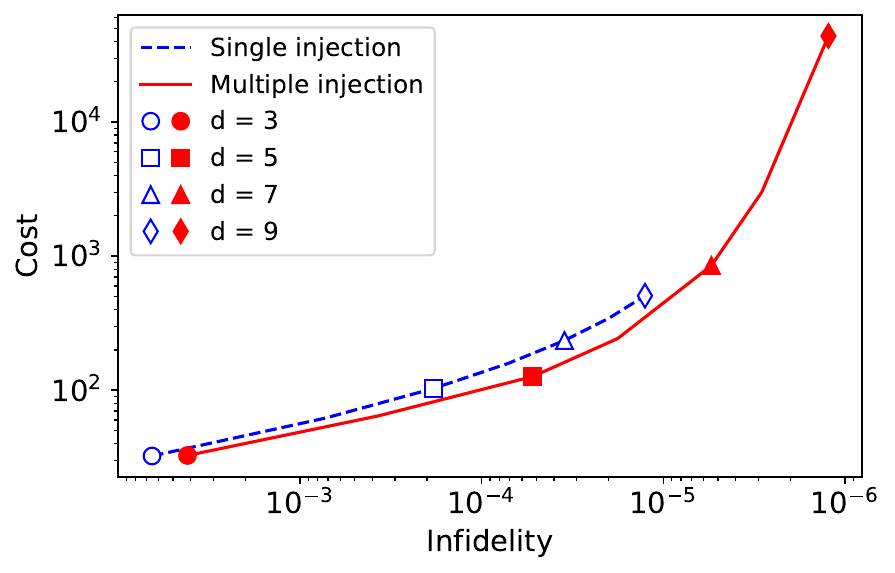}
\caption{Resource overhead comparison between single and multiple injection schemes for achieving different infidelities under a $Z$ error, the target logical state is $R_{Z,L}(\theta_{L})\ket{+}$ with $\theta_{L}=2\pi\times 10^{-3}$. The first left cycle dots of two protocol both label the situation for a distance 3 surface code, and the rest is similar. It is obvious multiple injection can achieve lower infidelities with smaller code distance $d$.}
\label{Infidelity with Cost}
\end{figure}

\subsection{The resource overhead}

The other advantage of multiple injection lies in the reduced number of physical qubits required to implement continuous fault tolerance, by changing the space-cost to time-cost.
Here, we compare the resource overhead of single and multiple injection protocols. Similarly to before, we set that all the physical qubits have the same physical rotation angle, and all the logical chains contribute the same logical rotation angle, for the demonstration purpose. We define the resource overhead as the space-time cost, the product of the number of qubits and the number of time steps that are occupied for computation. A time step is defined as one stabilizer measurement cycle, which encompasses the time required to implement a Hadamard circuit for measuring a stabilizer. Here, we assume that the stabilizers must be measured one at a time, through measurement circuits. This implies that we measure the first stabilizer and obtain its value, then proceed to the measurement circuit for the next stabilizer. In this scenario, the resource overhead is significantly reduced by immediately stopping and restarting the preparation if any stabilizer does not pass the post-selection. Moreover, if it passes the post-selection, we repeat the stabilizer measurement cycle again to detect whether an error occurred in the previous stabilizer measurement cycle.

For the detailed formulation of the resource overhead, we introduce the total expected numbers of repetition, $N = 1/P_{t}$, where $P_{t}$ labels the total probability for passing all the post-selection. For the single injection, $P_{t}$ is trivially equal to the probability $P_{t} = P_c$. And for multiple injection, $P_{t}$ equals to the probability for all the chains to pass the post-selection, i.e., $P_{t} =  (P_c)^{n}$, where $n$ labels the total number of operated chains. Thus the resource overhead is
\begin{equation} \label{resource overhead}
R_o=\sum_{i=1}^m i(d-1)\cdot N_s \cdot N \cdot P_{n}^{i}(1-P_{n})^{m-i},
\end{equation}
where $N_s$ labels the number of physical qubits corresponding to a stabilizer. Generally, $N_s = 2 (4)$ if the corresponding stabilizer has 2(4) physical qubits. 

Note that the resource overhead appears to increase exponentially with the number of injection chains, since the success probability of multiple injection is a simple product of the success probability for each individual chain. However, this will not significantly affect the success probability, especially for small rotation angles and middle-scale code distances. Considering a target logical rotation $R_{L}(\theta_L)$, it just need to achieve the angle $\theta_c=\theta_L/n$ for each operated chain in the multiple injection scheme. Thus, the success probability of each individual chain in the multiple injection scheme becomes larger compared to that of the single injection scheme. With higher success probability for each chain and fewer qubits, the multiple injection scheme still performs better in the middle scale of code distances, which is the state-of-the-art technology of leading candidates for physical implementing quantum computers. To show the detailed comparison between single and multiple injection scheme, we formulate the promised logical state infidelity with one error gate, and corresponding resource overhead. As shown in Fig. \ref{Infidelity with Cost}, for small rotation angles, the resource overhead of the multiple injection scheme is lower than that of the single injection scheme. And, the break-even logical rotation angle is about $\theta_{L}=\pi/50$. 
Nevertheless, this increase of the resource overhead just means more gate operations. That is, for a certain target of infidelity, the multiple injection requires fewer physical qubits. Therefore, we effectively transforming space costs into time costs, highly reducing the requirement for a larger code distance, i.e., a larger of qubit number.

\section{Conclusion}  
In conclusion, we introduce an efficient scheme that utilizes multiple logical chains with identical or varied logical rotation angles to initialize non-Clifford states. This approach mitigate the requirement of the code distance to maintain the continuous fault tolerance, which is currently a main limiting factor for scalable quantum computation. Therefore, our scheme ensures continuous fault tolerance in middle-scale qubit lattices with the surface code encoding.

\bigskip
The data that support the findings of this study are available from the authors upon reasonable request.


\acknowledgments
This work was supported by the National Natural Science Foundation of China (Grant No. 12275090), the Guangdong Provincial Quantum Science Strategic Initiative (Grant No. GDZX2203001), and the Innovation Program for Quantum Science and Technology (Grant No. 2021ZD0302303).

\appendix

\section{Logical operators}

In this section, we detail the logical operators for different distances $d$. Same as in the maintext, a surface code is firstly initialized into the Clifford state of $\ket{+}_{L}$ in  a fault-tolerant way. Then, we apply single-qubit rotation gates around the $Z$ axis to a set of qubits $Q$, i.e.,
\begin{eqnarray}
\label{Append_R_z to Q}
\prod_{j\in Q} R_{z}(\theta_{p_j})
=\prod_{j\in Q}(\cos{\frac{\theta_{p_j}}{2}}\cdot I+\text{i}\sin{\frac{\theta_{p_j}}{2}}\cdot Z),
\end{eqnarray}
where $\theta_{p_i}$ labels the angle of rotation around the $Z$ axis for the $i$th physical qubit. If the injected qubits form a logical chain, $C$, we obtain
\begin{equation}
\label{Append_R_z to Q 2}
(\prod_{j\in C}\cos{\frac{\theta_{p_j}}{2}}\ket{+}_L+\prod_{j\in C}\text{i}\sin{\frac{\theta_{p_j}}{2}}\ket{-}_L)+\text{others}.
\end{equation}
When $d$ is odd, we obtain a general logical Z rotation as it remains a imaginary unit $\text{i}$, we can write the logical operator as
\begin{equation} \label{Append_Logical operator_odd}
\prod_{j\in Q} R_{z}(\theta_{p_j})=\sqrt{P}R_{Z,L}(\theta_c)+\text{others}.
\end{equation}
where $Z_{L} = \prod_{j\in c} Z_{j}$, $R_{Z,L}$ labels the logical $Z$-rotation, and $\theta_{c}$ labels the rotation angle contributed by the logical chain to the final generated logical state.
When $d$ is even, we deal with the evolution operator, extract the imaginary unit $\text{i}$. The logical operator is,
\begin{equation}
\label{Append_Logical operator_even}
\prod_{j\in Q} R_{z}(\theta_{p_j})=\sqrt{P}e^{-i\frac{\pi}{4}}R_{X,L}(\frac{\pi}{4})R_{Z,L}(\theta_c)+
\text{others}.
\end{equation}

\section{Fidelity calculation utilizing one qubit}

For the calculation of infidelities, we employ single qubit states rather than logical encoded states to derive the infidelity. We calculate the logical rotation angle for both error case and no error case. Specifically, the final logical state, without errors, becomes
\begin{equation}
\label{Append_states}
R_{Z,L}(\theta_{L})\ket{+}_{L}=\cos{\frac{\theta_{L}}{2}}\ket{+}_{L} +\text{i}\sin{\frac{\theta_{L}}{2}}\ket{-}_{L},
\end{equation}
and the final logical state with a $Z$ error becomes,
\begin{equation}
\label{Append_states_Z error}
R_{Z,L}(\theta_{L,e})\ket{+}_{L}=\cos{\frac{\theta_{L,e}}{2}}\ket{+}_{L} +\text{i}\sin{\frac{\theta_{L,e}}{2}}\ket{-}_{L}.
\end{equation}
Generally, for calculating the fidelity, $\lvert\braket{\psi_{e} \lvert \psi}_L\lvert^2$, the inner products of the logical basis states is actually equal to products of the physical qubit states, yield $\braket{+ \lvert +}_L =\braket{+ \lvert +} = 1$ and $\braket{+ \lvert -}_L =\braket{+ \lvert -} = 0$. Given this consideration, it is reasonable to employ the states of a single qubit, $\ket{+}$ and $\ket{-}$, rather than logical state to compute infidelities, as the infidelity is solely dependent on the superposition coefficient.

\end{document}